\documentclass[12pt]{article} 
\usepackage{graphicx,floatflt,amssymb,epsf,rotate} 
\begin{document} 
\begin{flushright} 
HRI-P-10-01-002 \\
\end{flushright} 
 
\vskip 30pt 
 
\begin{center} 
{\Large \bf Type I and $new$ seesaw in left-right symmetric theories}\\
\vspace{0.5in}
{ {\bf Joydeep Chakrabortty
}  }\\ 
\vspace{0.2in}
   {\sl Harish-Chandra Research Institute,\\
Chhatnag Road, Jhunsi, Allahabad  211 019, India} \\
\normalsize 
\end{center} 

\vspace{0.5in}
\begin{abstract}
We extend the Type I seesaw and suggest a $new$ seesaw mechanism to generate neutrino masses 
within the left-right symmetric theories where parity is spontaneously broken. 
We construct a next to minimal left-right symmetric model where neutrino masses are determined irrespective 
of the B-L breaking scale and call it the $new$ seesaw mechanism. In this scenario B-L scale can be very low. 
This makes B-L gauge boson and the quasi-Dirac $heavy$ leptons very light. These TeV scale particles could 
have large impact on lepton flavor and CP violating processes. We also shed light on the phenomenological aspects 
of the model within the reach of the LHC.
\end{abstract}
\pagebreak

\renewcommand{\thesection}{\Roman{section}} 
\setcounter{footnote}{0} 
\renewcommand{\thefootnote}{\arabic{footnote}} 
\noindent

\section{Introduction}
The Large Hadron Collider (LHC) is expected to reveal the theory beyond the tera (TeV) scale. 
A most $popular$ question arises: 
Will it be possible to get some signatures of physics beyond the TeV scale? Physics in the neutrino sector 
might be a smoking gun to probe the high scale theories. The small but non-zero neutrino masses 
are unexplained within the SM. Experiments like SNO, Kamland, K2K and MINOS \cite{solar, kl, k2k, minos} provide information on the two mass squared differences $\Delta m_{21}^2$ and $\Delta m_{31}^2$ 
\footnote{ $\Delta m_{ij}^2=m_i^2-m_j^2$.} and on the two
mixing angles $\theta_{12}$ and $\theta_{23}$. The third mixing angle
$\theta_{13}$ is not yet determined. Null result of the CHOOZ \cite{chooz} experiment confirms
 this to be certainly small. The
current $3 \sigma$  allowed intervals  of the oscillation parameters are
given as \cite{limits}
\begin{equation}
 7.1 \times 10^{-5} \rm{eV^2}<\Delta m_{21}^2 < 8.3\times 10^{-5}
 \rm{eV^2},\hspace*{0.1cm} 2.0 \times 10^{-3} \rm{eV^2}<\Delta m_{31}^2 <
 2.8\times 10^{-3} \rm{eV^2}
\label{eq:delmasses}
\end{equation}
 \begin{equation}
   0.26< \sin^2 \theta_{12}<0.42,\hspace*{0.1cm}    0.34 <\sin^2
   \theta_{23}<0.67 , \hspace*{0.1cm}  \sin^2 \theta_{13}<0.05 ~.
\label{eq:angles}
\end{equation}

There are several models, beyond SM, which attempt to explain the origin of the neutrino masses. Seesaw mechanism where light neutrino masses are generated by integrating out the heavy 
particles is one of the popular and well established models. 
Studies are going on to explore different possibilities within the seesaw models, namely 
Type I, II, III, and Inverse seesaw depending on the nature of the $heavy$ particles. These $heavy$ particles carry the signatures of the high scale theories and their need strongly motivates to go beyond the SM. 

In case of Type I seesaw mechanism~\cite{TypeI}, one should have
at least two fermion singlets $\nu_R$ (right-handed neutrinos) 
and the neutrino masses read as $m_{\nu_L}^I \simeq m_D^2/ M_{\nu_R}$,
where $m_D$ is the Dirac mass term and $M_{\nu_R}$ are the 
Majorana (lepton number violating) masses of the right handed neutrinos. If $m_D$ $\approx$ 100 GeV and $M_{\nu_R} \approx 10^{13}$ GeV, one obtains the natural value for the neutrino masses $m_\nu \approx 1$ eV.

In Type II seesaw mechanism~\cite{TypeII} the SM is extended by 
an $SU(2)_L$ triplet Higgs $\Delta$. In this scenario the neutrino
masses are given by $m_{\nu_L}^{II} \simeq Y_\nu  v_{\Delta}$, where
$v_{\Delta}$ is the vacuum expectation value (vev) of the neutral component of the triplet 
and $Y_\nu$ is the Yukawa coupling. $v_{\Delta} \simeq \mu v^2/M_{\Delta}^2$, 
where $M_{\Delta}$ is the mass of the triplet and $\mu$ is
the trilinear coupling between the SM Higgs and the triplet. Light neutrino masses $m_\nu \approx 1$ eV 
if we assume $Y_\nu \approx 1$, $v \approx 100$ GeV and $\mu \sim M_{\Delta} \approx 10^{13-14}$ GeV.

For the Type III seesaw mechanism~\cite{TypeIII} one needs to add at least two extra matter
fields in the adjoint representation of $SU(2)_L$ with zero hypercharge
to generate neutrino masses, $m_{\nu_L}^{III} \simeq M_D^2/M_{\Sigma}$. 
Here $M_{\Sigma}$ stands for the mass of the fermionic triplets and $M_D$ is 
the Dirac coupling. 

Recently these seesaw mechanisms are widely investigated in the context of 
$SU(5)$, $SO(10)$, and $E(6)$ grand unified theories (GUT) and left-right symmetric theories 
\cite{recentlr}. In minimal $SU(5)$ theory neutrinos are massless. If we add 24-dimensional adjoint fermions, neutrino masses can be generated through Type I and Type III seesaw mechanism \cite {Perez}. 
In $SO(10)$ fermions are 16-dimensional and has a room for right handed neutrinos. Thus Type I seesaw is 
a natural outcome in minimal $SO(10)$ models. Type II and Type III seesaw mechanisms can be achieved by 
adding 126-dimensional Higgs and 45-dimensional adjoint fermions respectively. $E(6)$ accumulates the features of $SO(10)$ as it contains $SO(10)\otimes U(1)_X$ as a maximal subgroup. Fermions are in 27-dimensional irreducible representation that contains an $SO(10)$ singlet. 
In the $new$ seesaw mechanism this singlet plays an important role.

Other possible solutions are implemented to resolve neutrino mass problem within the framework 
of heterotic superstring models \cite{superstring} or $n_G$ generation models \cite{ng-generation}. 
It has been shown in \cite{pilaftsis} that even if the total lepton-number is conserved 
the separate lepton-numbers are violated in general. The decays of charged leptons 
leading to (individual) lepton-number and lepton-flavour violations are discussed \cite{Ilakovac} 
in the context of a seesaw-type extensions of the SM with left handed and/or right handed weak 
isosinglets. The charged leptons decay mainly as: 

(i) $\ell \rightarrow \ell' \gamma$ :

The experimental bounds arising from these type of decays are~\cite{PDG}
\begin{equation}
 B(\tau\to e\gamma)< 1.1\times 10^{-7},\quad B(\tau \to \mu\gamma) < 6.8\times 10^{-8},
\quad B(\mu\to e \gamma) < 1.2\times 10^{-11}.
\end{equation}
The photonic decay modes are extremely suppressed. Thus it is very hard to probe 
the heavy neutrino physics using these decays.

(ii) $\ell \rightarrow \ell' \ell_1 \bar{\ell}_2$:

 In three generation model the decaying charged lepton might be either $\tau$ or $\mu$. \\
The possible three body decay modes of $\tau$ \cite{tau-decay}:
\begin{eqnarray}
   \tau^- &\to \mu^-\mu^-\mu^+ , \nonumber\\
          &\to e^-e^-e^+,\nonumber\\
           &\to e^-\mu^-\mu^+ , \nonumber\\
            &\to \mu^-e^-e^+   \nonumber
\end{eqnarray}
- these are lepton-flavour violating and
\begin{eqnarray}
          &\to e^+\mu^-\mu^-, \nonumber\\
     &\to \mu^+ e^-e^- \nonumber
 \end{eqnarray}
- these are lepton-number violating decays. There could be other lepton-number 
and lepton-flavour violating decyas of $\tau$ that include hadrons in the final states:
\begin{eqnarray}
   \tau^- &\to e^-\pi^+\pi^-,~~ e^- \rho^0 , \nonumber\\
         &\to \mu^-\pi^+\pi^-, ~~\mu^- \rho^0 , \nonumber\\
           &\to e^-\pi^+K^-, ~~e^- K^{*0} , \nonumber\\
          &\to \mu^-\pi^+K^-, ~~\mu^- K^{*0} , \nonumber
\end{eqnarray}
- lepton-flavour violating and
\begin{eqnarray}
        \tau^- &\to e^+\pi^-\pi^-, \nonumber\\
         &\to \mu^+\pi^-\pi^-, \nonumber\\
           &\to e^+\pi^-K^-, \nonumber\\
          &\to \mu^+\pi^-K^-, \nonumber
 \end{eqnarray}
- lepton-number violating decays.

The decays The branching ratios of $\tau$ decays into three charged leptons can be large $\sim$ $10^{-6}$.

$SO(10)$ contains $SU(2)_L \otimes SU(2)_R \otimes SU(3)_C \otimes U(1)_{B-L}$ as 
a maximal subgroup which is left-right symmetric. We start with left-right symmetric theories and discuss two possible mechanisms for neutrino mass generation. In section II, we propose a model where right handed neutrino masses are generated through a dimension-5 operator, like $LLHH/M$, and light neutrino masses are the outcome of a Type I seesaw. 
In section III, we extend our model by two Higgs doublets and a singlet fermion. Here neutrino masses are generated through a $new$ type of seesaw mechanism where light neutrino masses are independent of the B-L scale.
We discuss phenomenological implications of these models.

\section{Type-I seesaw in left-right theories}
The so-called left-right symmetric models are one of the 
most appealing extensions of the SM where one can understand 
the origin of parity violation in a simple way and we can generate
neutrino masses. The simplest theories are based on the gauge group
$ SU(2)_L \otimes SU(2)_R \otimes SU(3)_C \otimes U(1)_{B-L}$.
Here $B$ and $L$ stand for baryon and lepton number, respectively.
The matter multiplets for quarks and leptons are given by
\begin{equation}
Q_L = \left(
\begin{array} {c}
u_L \\ d_L
\end{array}
\right) \ \equiv \ (2,1,3,1/3),~
Q_R = \left(
\begin{array} {c}
 u_R \\ d_R
\end{array}
\right) \ \equiv \ (1,2,3,1/3),
\end{equation}
\begin{equation}
l_L = \left(
\begin{array} {c}
 \nu_L \\ e_L
\end{array}
\right) \ \equiv \ (2,1,1,-1),~
l_R = \left(
\begin{array} {c}
 \nu_R \\ e_R
\end{array}
\right) \ \equiv \ (1,2,1,-1).
\end{equation}

In our model we have two Higgs doublets:
\begin{equation}
H_L = \left(
\begin{array} {c}
h_{L}^{+} \\ h_{L}^{0}
\end{array}
\right) \ \equiv \ (2,1,1,1),~
H_R = \left(
\begin{array} {c}
 h_{R}^{+} \\ h_{R}^{0}
\end{array}
\right) \ \equiv \ (1,2,1,1)
\end{equation}
and a bidoublet Higgs:
\begin{equation}
\Phi = \left(
\begin{array} {cc}
 \phi_1^0   &  \phi^+_2 \\
 \phi^-_1  & \phi_2^0
\end{array}
\right) \equiv (2, 2, 1, 0) ~~~ \mbox{and}~~~ \tilde{\Phi}= \sigma_2 \Phi^* \sigma_2.
\end{equation}
Under the left-right parity transformation one has the following relations
\begin{equation}
F_L \longleftrightarrow F_{R}
\end{equation}
where $F$ = $Q$, $l$, $H$.

The relevant Yukawa interactions for quarks in this 
context are given by
\begin{eqnarray}
{\cal L}_Y^{quarks} &=& \bar{Q}_L( Y_1 \Phi  +  Y_2 \tilde{\Phi}) Q_R  + h.c.
\end{eqnarray}
Once the bidoublet gets the vev the quark mass matrices read as
\begin{eqnarray}
M_U & = & Y_1 v_1 \ + \ Y_2 v_2^*~~~ \mbox{and} ~~~
M_D =  Y_1 v_2 \ + \ Y_2 v_1^*,
\end{eqnarray}
with $v_1= \langle \phi_1^0 \rangle$ and $v_2 = \langle \phi_2^0 \rangle$.
In the case of the bidoublet one has the following transformation under the left-right parity
\begin{equation}
\Phi \longleftrightarrow \Phi^\dagger,
\end{equation}
and $Y_1 = Y_1^\dagger$ and $Y_2 = Y_2^\dagger$.

In this context the charged lepton masses are generated through the interactions
\begin{equation}
{\cal L}_l= \bar{l}_L ( Y_3  \Phi  +  Y_4 \tilde{\Phi}) l_R  + h.c. 
\end{equation}
and the relevant mass matrix is given by
\begin{equation}
M_e= Y_3  v_2  + Y_4 v_1^*.
\end{equation}
At the same time Dirac mass matrix for the neutrinos is written as:
\begin{equation}
m_\nu^D= Y_3 v_1  +  Y_4 v_2^*.
\end{equation}
However, in this case one has the same situation as in the SM plus
right-handed neutrinos where we can assume a small Dirac Yukawa coupling
for neutrinos. In this model right handed neutrino can not have any tree level mass. 
Once the $H_L$ and $H_R$ acquire the vevs, $v_L$ and $v_R$ respectively, left-right symmetry 
is broken. In our model $v_L$=0 and the vev ($v_R$) of $H_R$ sets the scale where 
$SU(2)_R\otimes U(1)_{B-L}$ {\footnote {We call it B-L (breaking) scale in our discussion.}}
is broken to $U(1)_Y$. 
This vev generates the right-handed neutrino Majorana mass term
\begin{equation}
M_{\nu_R} \approx \frac{\eta v_R^2}{M}
\end{equation}
through a non-renormalisable operator, like $\eta \ell \ell H H /M$, where $\eta<1$ is the strength of the 
non-renormalisable coupling and $M$ is the high scale which may be around unification scale.
The neutrino mass matrix in the basis ($\nu_{L},\nu_R$) is :
\begin{eqnarray}
\label{mnu}
M_\nu = \pmatrix{
  0 & m_{\nu}^D 
\cr
  (m_\nu^D)^{T}&  M_{\nu_R}
\cr}\;.
\end{eqnarray}
Now assuming $M_{\nu_R} >>~ m_D$, we have light neutrino masses:
\begin{equation}
m_{\nu_{L}}=m_\nu^D M_{\nu_{R}}^{-1} (m_\nu^D)^{T}
\end{equation}
As we have seen in this case though one has a single seesaw mechanism we
can have an interesting scenario for the LHC where the right handed neutrinos
are at the scale, $M_{\nu_R} \approx 1$ TeV. Therefore, as it is well known
in this case one gets small neutrino masses, $m_\nu \approx 1$ eV, if the Yukawa 
couplings are very small. If we assume $M$ $\sim 10^{19-17}$ GeV, 
the B-L scale will be, $v_R \sim 10^{11-10}$ GeV to get the consistent
right handed neutrino mass term. The nice feature of this model is having TeV scale right handed neutrinos 
quite naturally without lowering the B-L scale. This possibility might be testable at collider experiments.
Because of the low right handed neutrino mass, signatures of the $\nu_R$ can be grabbed at the LHC 
\cite{seesawlhc}. The main production channels of the single heavy right handed neutrino are:
\begin{eqnarray}
p p  & \rightarrow & q \bar{q'} \rightarrow W^* \rightarrow \ell^\pm \nu_R, \nonumber \\
    &  \rightarrow & q \bar{q} \rightarrow Z^* \rightarrow \nu \nu_R, \nonumber \\
    &  \rightarrow & gg \rightarrow H^* \rightarrow \nu \nu_R, \nonumber
\end{eqnarray}
and the pair production signal is:
\begin{equation}
 p p \rightarrow q \bar{q} \rightarrow Z^* \rightarrow \nu_R \nu_R. \nonumber
\end{equation}
The production cross sections are very much suppressed by the mixing matrix elements ($V_{l\nu}$) 
and the SM backgrounds. \\
The decay widths of the heavy right handed Majorana fermions are: 
\begin{eqnarray}
\Gamma(\nu_R \to \nu_l H ) & =& \frac{g^2}{64\pi} |V_{\ell\nu_R}|^2 \frac{M_{\nu_R}^3}{M_W^2}
(1- \frac{m_H^2}{M_{\nu_R}^2})^{2} ,\nonumber \\
\Gamma(\nu_R \to \nu_l Z ) &=& \frac{g^2}{64\pi c_W^2} |V_{\ell \nu_R}|^2 \frac{M_{\nu_R}^3}{M_Z^2}
(1- \frac{m_Z^2}{M_{\nu_R}^2}) (1+\frac{m_Z^2}{M_{\nu_R}^2}-2\frac{m_Z^4}{M_{\nu_R}^4}) ,\nonumber \\
\Gamma(\nu_R \to  \ell^\mp W^\pm ) &=& \frac{g^2}{64\pi} |V_{\ell \nu_R}|^2 \frac{M_{\nu_R}^3}{M_W^2}
(1- \frac{m_W^2}{M_{\nu_R}^2}) (1+\frac{m_W^2}{M_{\nu_R}^2}-2\frac{m_W^4}{M_{\nu_R}^4}) ,\nonumber
\end{eqnarray}
where $V_{\ell \nu_R}$ is the small mixing, $|V_{\ell \nu_R}|^2$ $\sim$ $10^{-3}-10^{-4}$.\\
There is a lepton number violating process
\begin{equation}
\nu_R\nu_R \rightarrow l^{\pm} l^{\pm} W^{\mp}W^{\mp} ~~~(l=e, \mu, \tau) 
\end{equation}
 - Majorana signal.
If we consider that $W'$s decay hadronically then in the final state we will have same sign dileptons 
($SSD$) and 4 $jets$. The overall branching fraction for this is 
\begin{equation}
BR(\nu_{R_{i}} \nu_{R_{i}} \rightarrow l^{\pm} l^{\pm}~ +~4 ~jets) \approx  \frac{1}{18}.
\end{equation}
There could be other channels like $l^{\pm} l^{\mp}+ ~4~ jets$, but these are not 
lepton number violating processes. 
We can also look at the leptonic decay of one of the $W'$s 
\begin{equation}
\nu_{R_{i}} \nu_{R_{j}} \rightarrow l^{\pm} l^{\pm} W^{\mp}W^{\mp}\rightarrow l^{\pm} 
l^{\pm}l^{\mp}~+~ MET+~ 2 ~jets. 
\end{equation}
This might be a faithful decay mode to reconstruct the mass of the right handed neutrino 
as there is a single source of missing transverse energy (MET). The combinatorial background can be reduced by minimizing the difference of the masses of two right handed neutrinos.

We can have another scenario where $m_D \sim 100$ GeV, $v_R \sim 10^{15}$ GeV, $M \sim 10^{17}$ GeV 
and one finds $m_\nu \sim 1$ eV.
The scales, involved here are very high. This can be demonstrated 
as a low energy effect of some high scale theory, like GUT.

Type I seesaw mechanism for neutrino masses in the context of left-right symmetric models 
where parity is broken spontaneously has been discussed. We prescribe a left-right symmetric model 
where the light neutrino masses are generated by integrating out the heavy right handed neutrino through a Type I 
seesaw mechanism. We stick to the minimal matter fields in our model and have generated right handed neutrino masses
through a non-renormalisable operator. We point out the phenomenological rich and accessible aspects 
when $\nu_R$ is light. In the scenario where $\nu_R$ is very heavy and non-degenerate, its decay 
can lead to Type I leptogenesis.

\section{$new$  seesaw in left-right symmetric theories}
In this section we propose a $new$ type of seesaw realisation in the context of a non-supersymmetric 
left-right symmetric model where parity is spontaneously broken. We add an extra fermion 
which is singlet, $S$, under $ SU(2)_L \otimes SU(2)_R \otimes SU(3)_C \otimes U(1)_{B-L}$, but has a 
non-zero $U(1)_X$ charge. Thus the mass term, $\mu$, of this singlet 
is not allowed because of this $U(1)_X$ invariance. $\mu \rightarrow 0$ enhances the symmetry of 
the theory, as global lepton number symmetry is exactly conserved.
In $E(6)$ fermions are 27 dimensional and under $SO(10)\otimes U(1)_X$ decomposition reads as:
\begin{equation}
 27_F \equiv (1,4)_F \oplus (16,1)_F \oplus (10,-2)_F
\end{equation}
We consider $S$ to be an $SO(10)$ singlet field $(1,4)_F$ 
present in $27_F$ of $E(6)$. The $U(1)_X$ charge of this singlet prevents to 
have its mass term in the Lagrangian. Thus we put $\mu = 0$ in our model. 
The other fermions $(10,-2)_F$ do not mix and they are heavy. 
The field contents of this model are same as in the previous one with an added extra singlet fermion 
$(1,1,1,0)$ and two more Higgs doublets: $H_L^*\equiv (2,1,1,-1)$, $H_R^*\equiv(1,2,1,-1)$ 
under left-right symmetric gauge group. Here only $H_R^*$ gets vev, $v_R^*=v_R$ and $v_L^*=0$. 
We need to introduce $H_L^*$ to accommodate this in $16_H \oplus \overline{16}_H$ of $SO(10)$.\\
Now we can have extra Dirac couplings in the Lagrangian. 
When the Higgses get vevs
\begin{equation}
Y \bar{l}_R S H_R^*  ~+~ h.c.
\end{equation}
and
\begin{equation}
 \frac{1}{M} \bar{l}_L S (Y_5 ~\Phi + Y_6~ \tilde{\Phi})~ H_R^* ~+~ h.c.
\end{equation}
generate $\nu_R -S$ and $\nu_L -S$ couplings respectively.\\
The neutrino mass matrix in the basis ($\nu_{L}, \nu_R$, $S$) reads as:
\begin{eqnarray}
\label{mnu}
M_\nu = \pmatrix{
  0 & m_\nu^D &  M_D v_R/M
\crcr
  (m_\nu^D)^{T}&  \eta v_R^2/M & Y v_R
\crcr
M_D^T v_R/M & Y^T v_R  & 0
\crcr}\;.
\end{eqnarray}
where $M_D=(Y_5 v_1 + Y_6 v_2)$. In this model we consider $\eta$ to be very small such that 
$M_{\nu_{R}} \rightarrow 0$. After diagonalizing the above matrix, it can be seen that the scale $v_R$ 
drops out from the light neutrino masses:
\begin{eqnarray}
 m_{\nu_{L}}& = &(m_\nu^D ~~ M_D v_R/M) \pmatrix{
 0 & Y v_R
\crcr
 Y^T v_R  & 0
\crcr}^{-1} (m_\nu^D ~~ M_D v_R/M)^T \\ \nonumber
&=&\frac{1}{M} [m_\nu^D(M_D Y^{-1})^T+(M_D Y^{-1}) (m_\nu^D)^T].
\end{eqnarray}
$m_{\nu_L}$ is suppressed by $M$ and independent of the B-L scale. Thus, the B-L scale 
can be made low without affecting the small neutrino masses. The B-L gauge boson can be at TeV scale 
and possibly produced at the LHC. The $heavy$ neutrinos that are involved in 
seesaw mechanism having masses at $v_R$ scale are light now. This has rich phenomenological implications. 
Exchange of these particles generates lepton-flavour and (individual) lepton-number 
violating processes, say $\ell \rightarrow \ell' \gamma$ and 
$\ell \rightarrow \ell' \ell_1 \bar{\ell}_2$, at large rates as in 
the inverse seesaw models \cite{Ilakovac,branco}. This model can 
be tested at colliders and in flavor sector. 
If the right handed neutrino is lighter than $S$ then we could have Type I leptogenesis 
with extra vertex contributions containing $S$ in the loop. \\
This model is also testable at the LHC. The TeV scale right handed neutrinos co-exist with the prediction of small neutrino masses without fine tuning the couplings. This is a nice feature of this model. Thus the decays of TeV scale right handed neutrinos, discussed in the earlier section, fit naturally here. 
The form of the mass matrix is same as in reference \cite{Malinsky} but the mechanism and the elements are different.

\vskip 20pt

{\bf Acknowledgements:} I would like to thank Srubabati Goswami, Bruce Mellado, Biswarup Mukhopadhyaya, 
and Amitava Raychaudhuri for many useful suggestions. This research has been supported by funds from the XIth Plan `Neutrino physics' and RECAPP projects at HRI.

\end{document}